\documentclass{aa}  
\usepackage{graphicx}
\usepackage{txfonts}
\usepackage{natbib}
\usepackage{color}
\usepackage{hyperref}
\hypersetup{colorlinks=True,citecolor=blue,allcolors=blue}
\usepackage{tabularx}
\usepackage{tabu}
\usepackage{float}
\usepackage{xcolor}
\pdfoutput=1 

\begin{document} 

   \title{Continuum reverberation mapping of MCG 08-11-011\thanks{Light curves are only available at the CDS via anonymous ftp to cdsarc.u-strasbg.fr (xxx.xx.xxx.x) or via http://cdsarc.u-strasbg.fr/viz-bin/cat/J/A+A/xxx/xxx.}}
   
\author{C. Fian\inst{1,2,3,4}, D. Chelouche\inst{3,4}, S. Kaspi\inst{2},  C. Sobrino Figaredo\inst{3,4}, T. Lewis\inst{5}, S. Catalan\inst{4}}
\institute{Departamento de Astronom\'{i}a y Astrof\'{i}sica, Universidad de Valencia, E-46100 Burjassot, Valencia, Spain; \email{carina.fian@uv.es} \and School of Physics and Astronomy and Wise Observatory, Raymond and Beverly Sackler Faculty of Exact Sciences, Tel-Aviv University, Tel-Aviv, Israel \and Department of Physics, Faculty of Natural Sciences, University of Haifa,
Haifa 3498838, Israel \and Haifa Research Center for Theoretical Physics and Astrophysics, University of Haifa,
Haifa 3498838, Israel \and NASA Postdoctoral Program Fellow, NASA Goddard Space Flight Center, Code 661, 8800 Greenbelt Rd, Greenbelt, MD 20771, USA}
 

  \abstract
  {}
   {We report the results from a photometric reverberation mapping campaign carried out with the C18 telescope at the Wise Observatory from 2019 to 2020, targeting the active galactic nucleus (AGN) MCG 08-11-011. The monitoring was conducted on a daily basis with specially designed narrow-band filters, spanning from optical to near-infrared wavelengths ($\sim4000$ to $8000$\AA) and avoiding prominent broad emission lines. We aim to measure inter-band continuum time lags, determine the size-wavelength relation, and estimate the host-subtracted AGN luminosity for this system.}
   {We used the point-spread function photometry to extract the continuum light curves and measure the inter-band time lags using several methods, including the interpolated cross-correlation function, the z-transformed discrete correlation function, a von Neumann estimator, JAVELIN (in spectroscopic and photometric mode), MICA, and a multivariate correlation function.}
   {We find wavelength-dependent lags, $\tau(\lambda)$, up to $\sim$7 days between the multiband light curves of MCG 08-11-011. The observed lags are larger than predictions based on standard thin-disk theory by a factor of $\sim3-7$. We discern a significantly steeper ($\tau \propto \lambda^{4.74}$) size-wavelength relation than the $\tau \propto \lambda^{4/3}$ expected for a geometrically thin and optically thick accretion disk, which may result from the contribution of diffuse continuum emission to the flux. These results are similar to those found by previous continuum reverberation mapping campaigns.}
   {}

\keywords{Accretion, accretion disks --- galaxies: active --- galaxies: Seyfert --- galaxies: nuclei --- galaxies: photometry --- galaxies: individual (MCG 08-11-011)}

\titlerunning{Continuum reverberation mapping of MCG 08-11-011}
\authorrunning{Fian et al.} 
\maketitle

\section{Introduction} 
Active galactic nuclei (AGNs) are among the most luminous sources of radiation in the Universe, and understanding their interior structure has been one of the major goals of extragalactic astrophysics. The current picture of the schematic sub-parsec-scale structure of an AGN includes three main components: a hot, X-ray emitting corona; an accretion disk around a supermassive black hole (SMBH), and a broad-line region (BLR) consisting of fast-orbiting photoionized gas and clouds. On scales of parsec to hundreds of parsecs, the AGN consists of an obscuring dusty torus and a narrow-line region (NLR) comprised of small, low-density gas clouds moving at lower velocities. Gravitational potential energy and viscous heating is converted into heat and radiation by the accretion of matter onto the central SMBH (e.g., \citealt{Page1974,Rees1984,Balbus1998}). The accretion disk thereby reaches temperatures of $10^5-10^6$K at its inner edge with a gradient to cooler temperatures at larger radii, leading to a continuum emission spectrum spanning from the extreme ultraviolet (UV) to the infrared (IR). The hottest parts of the accretion disk provide the ionizing photons that cause Doppler-broadened emission lines in the BLRs and NLRs (\citealt{Davidson1979,Veilleux1987}). Although this basic picture can explain most of the observational properties of AGNs (\citealt{Burbidge1967,Weedman1977,Shields1978,Elvis1994,Telfer2002}), the detailed geometry and kinematics of the interior structure remain poorly understood. Since the sub-parsec-scale structures are unresolved in even the closest AGN, information must be obtained by indirect means.

Reverberation mapping (RM; \citealt{Bahcall1972,Blandford1982,Peterson1993,Peterson2014}) is a powerful tool to probe compact structures in the central parts of AGNs. The basic principle of RM is to search for temporal correlations between the time-variable flux signals (intrinsic variability) and their light echoes at different wavelengths. Combined with the speed of light, the lag between those light echoes determines the characteristic size of the reverberation structure in the AGN. For example, gas in the outer part of the accretion disk reprocesses (as variable optical flux) variations emitted in the far and extreme UV by the inner parts of the accretion disk. Measuring the lag between the primary UV signal and light echoes at longer wavelengths provides an estimate of the accretion disk's spatial extent. Recent findings suggest that the disk sizes are larger than the predictions from standard models (e.g., \citealt{Fian2022,Fausnaugh2017,Fausnaugh2018,Edelson2015,Edelson2019}). Accretion disk sizes considerably larger than predicted by theory have also been found in microlensing studies of gravitationally lensed quasars (e.g., \citealt{Morgan2010,Blackburne2011,Fian2016,Fian2018,Fian2021,Cornachione2020b,Cornachione2020a}). In addition, continuum time lags across the accretion disk provide information about the disk's temperature gradient, and it appears that they are flatter than expected (\citealt{Motta2017,Cornachione2020c,Jimenez2014}).

Measuring inter-band continuum lags is extremely challenging, because the predicted size of accretion disks is only about one light day and monitoring campaigns require comparable or better cadence (i.e., on the order of one day or less) on timescales of weeks to months to resolve such short lags. In this work, we analyzed six months of densely sampled (daily cadence) photometric monitoring data of the Seyfert 1 galaxy MCG 08-11-011, and we present detections of optical and near-IR inter-band continuum time lags. In Section \ref{2}, we discuss our observations, the data reduction, and the light curves taken in multiple photometry bands. In Section \ref{3}, we describe our time series analysis and compare several tools to measure the inter-band continuum time lags. The results are presented in Section \ref{4}, including the time delays, the lag spectrum, the host-subtracted AGN luminosity, and a comparison with the theoretical disk sizes. Finally, we discuss and conclude our findings in Section \ref{5}.

\section{Observations and data reduction}\label{2}
\begin{figure}[h]
\centering
\includegraphics[width=9cm]{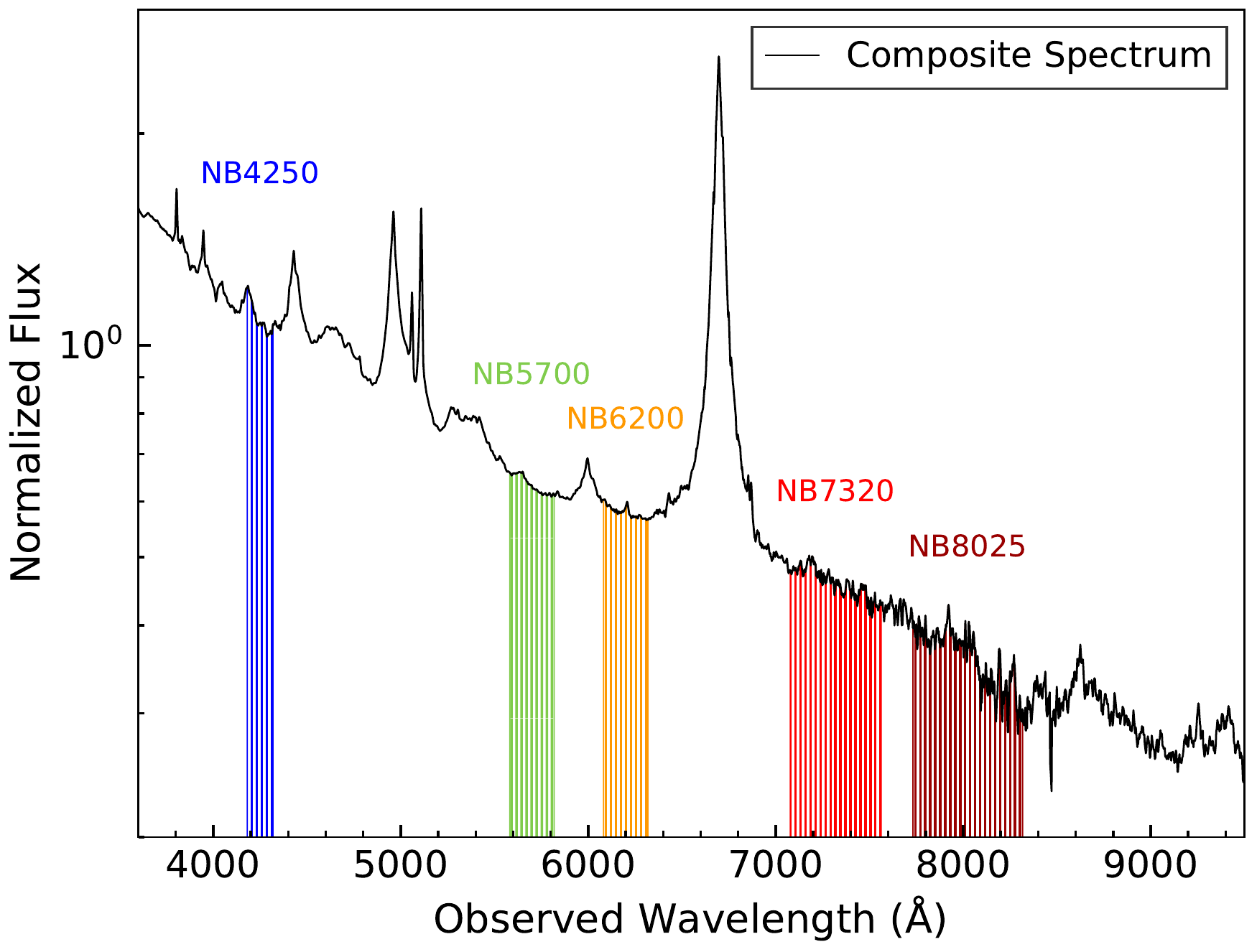}
\caption{Composite AGN spectrum of \citet{Glikman2006} shown at the quasar's redshift of $z = 0.0205$. The five emission-line-free NB passes of the filters which mainly trace the AGN continuum variations (and are used for the photometric monitoring presented in this work) are shaded in different colors (blue for NB4250, green for NB5700, orange for NB6200, red for NB7320, and dark red for NB8025).}
\label{spectrum}
\end{figure}

The ground-based photometric monitoring was conducted between October 2019 and April 2020 with the robotic C18 telescope (\citealt{Brosch2008}) of the Wise Observatory located in the Negev desert in southern Israel. We used the QSI 683 CCD (image sensor KAF-8300), which has $3326\times2504$ pixels of 5.4 $\mu$m in size. The pixel scale is 0.882 arcsec\,pix$^{-1}$, which gives a field of view of $48.9\times36.8$ arcmin ($0.815\times0.613$ degrees, corresponding to an area of 0.5 deg$^2$). The observations were carried out on a daily basis ($\sim$4 exposures per night in each filter) for a duration of almost six months. To trace the AGN continuum variations free of emission lines at the object's rest frame (MCG 08-11-011 is at a redshift of $z\sim0.0205$\footnote{https://ned.ipac.caltech.edu/}), five relatively narrow bands (NBs) at 4250, 5700, 6200, 7320, and 8025\AA\ were carefully chosen. In Table \ref{characteristics} we list the object's characteristics. In Table \ref{obs_fil} we summarize the filter and observation information, and Figure \ref{spectrum} shows the position of the NB filters together with the quasars' composite spectrum of \citet{Glikman2006}. The images were reduced following standard procedures performed with IRAF (including bias subtraction, dark current correction, and flat fielding for each filter), and we used the traditional point-spread function (PSF) fitting photometry to obtain the light curves. 
\begin{table}[h]
        \tabcolsep=0cm  
        \renewcommand{\arraystretch}{1.2}
        \caption{MCG 08-11-011 characteristics.}
        \begin{tabu} to 0.493\textwidth {X[c]X[c]X[c]X[c]X[c]} 

                \hline
                \hline 
                R.A. (J2000.0) & Dec. (J2000.0) & $m_V$ & $z$ & $\lambda L_\lambda$ (5100\AA)\\  
                (1) & (2) & (3) & (4) & (5)\\ \hline
                $05\ 54\ 53.6$ & $+46\ 26\ 21.6$ & $14.6$ & $0.0205$ & $4.21\pm0.65$ \\ \hline 
        \end{tabu}
    
        \vspace*{2mm}        
                \small NOTES. --- Cols. (1)--(2): Right ascension and declination from NED. Units of right ascension are hours, minutes, and seconds; units of declination are degrees, arcminutes, and arcseconds. Col. (3): $V$ magnitude from the \citet{Veron-Cetty2010} catalog. Col. (4): Redshift from the NASA/IPAC Extragalactic Database$^1$. Col. (5): Host-subtracted AGN luminosity interpolated to restframe 5100\AA\ (in units of $10^{43}$ ergs s$^{-1}$; see Section \ref{host} for a detailed discussion).\\
\label{characteristics} 
\end{table}

\begin{table}[h]
    \tabcolsep=0cm
        \renewcommand{\arraystretch}{1.2}
        \caption{Filter and observation information.}
        \begin{tabu}{X[c]X[c]X[c]X[c]X[c]} 

                \hline
                \hline 
                Filter & CWL (\AA) & $\lambda$ (\AA) & $\Delta \lambda$ (\AA) & $t_{exp}$ (s)\\  
                (1) & (2) & (3) & (4) & (5) \\ \hline
                NB4250 & 4250 & $4150-4330$ & $138$ & $350$ \\
                NB5700 & 5700 & $5550-5830$ & $238$ & $90$ \\
                NB6200 & 6200 & $6050-6350$ & $238$ & $110$ \\
                NB7320 & 7320 & $7040-7610$ & $488$ & $90$ \\
                NB8025 & 8025 & $7700-8370$ & $588$ & $100$ \\\hline 
        \end{tabu}
                
                \small \vspace*{2mm}NOTES. --- Col. (1): NB filter. Cols. (2)--(4): Central wavelength, minimum, and maximum wavelength at 0.01\% transmission, and FWHM in the units of \AA. Col. (5): Exposure time of each image in seconds.
\label{obs_fil} 
\end{table}

\begin{figure}
\centering
\includegraphics[width=9.1cm]{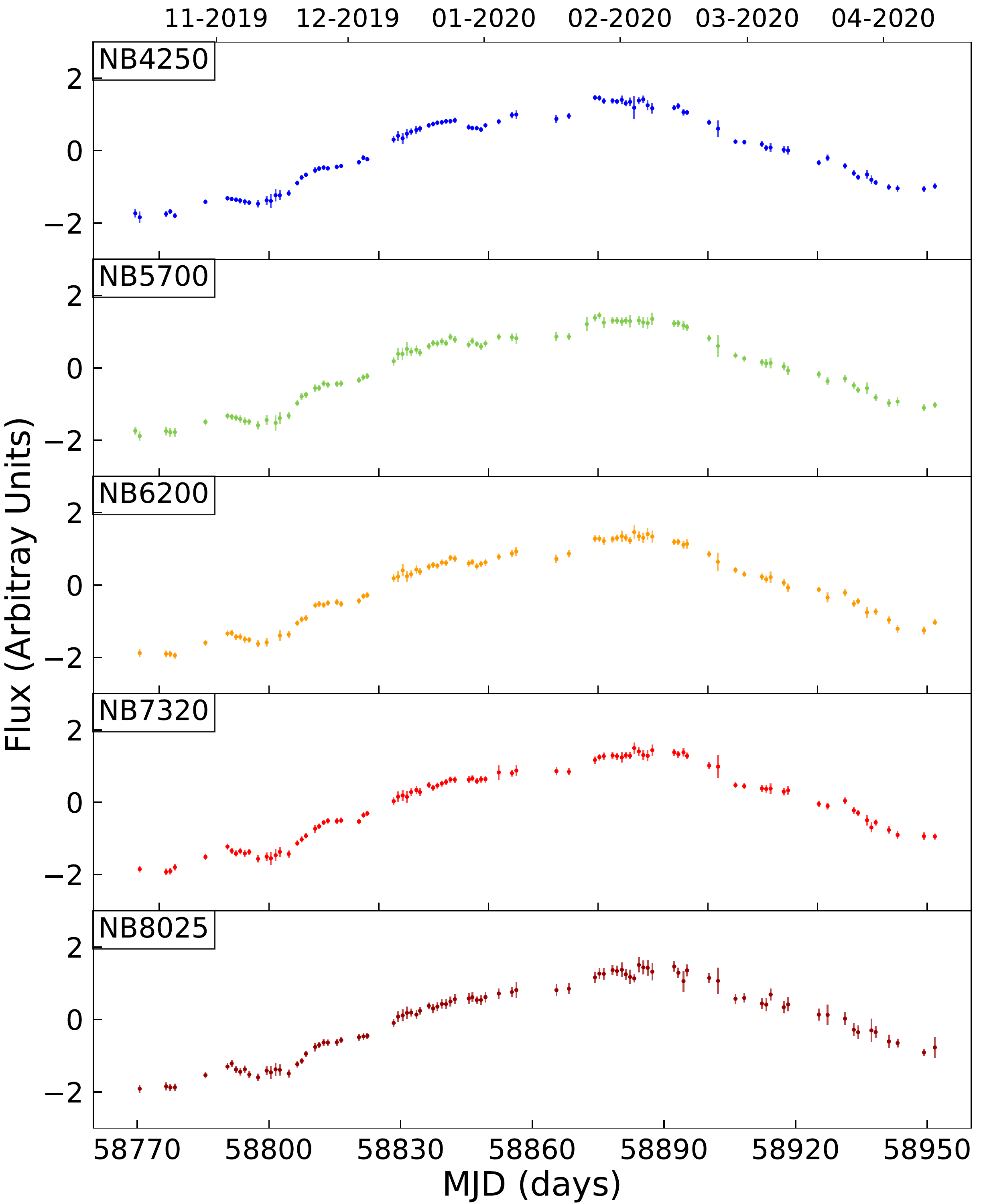}
\caption{PSF photometry light curves of the AGN continuum at 4250, 5700, 6200, 7320, and 8025\AA\ (from top to bottom) for the period between October 2019 and April 2020. The light curves are normalized to zero mean and unit standard deviation, and the fluxes are in arbitrary units.}
\label{lightcurves}
\end{figure}

We used the DAOPHOT (\citealt{Stetson1987}) package as implemented in IRAF and DAOSTAT (\citealt{Netzer1996}) to measure the magnitude of the objects in the images and to compute the light curves of the Seyfert 1 galaxy (see \citealt{Fian2022} for a detailed description). To obtain accurate measures for the magnitude at a given epoch, we discarded problematic exposures (due to low S/N, condensation rings on the CCD, and/or elongated stars caused by telescope tracking or auto-guider issues) from each night. After comparing consecutive points and removing points above a certain threshold, we are left with a set of good measurements per night (only one night has been discarded). Finally, we averaged the outlier-free exposures for each night, resulting in high S/N light curves consisting of $\sim$90 data points. In Figure \ref{lightcurves} we show the normalized-to-mean and unit standard deviation light curves for the different bands, and in Table \ref{var} we present the variability measures for each light curve. 

\footnotetext[2]{https://www.astromatic.net/software/swarp/}

\begin{table*}[h]
        \tabcolsep=0cm
        \renewcommand{\arraystretch}{1.2}
        \caption{Variability measures for the host-subtracted PSF photometry light curves.}
        \begin{tabu} to 1\textwidth {X[c]X[c]X[c]X[c]X[c]X[c]X[c]X[c]} 
                
                \hline
                \hline 
                Filter & $\overline{m}$ & $rms$ & $\delta$ & $\chi^2_\nu$ & $\sigma_N$ & $F_{var}$ & $Err\, F_{var}$ \\  
                 (1) & (2) & (3) &(4) & (5) & (6) & (7) & (8)\\ \hline
                NB4250 & $8.48$ & $2.25$ & $0.21$ & $139$ & $26.4$ & $0.264$ & $0.020$ \\
                NB5700 & $7.41$ & $1.91$ & $0.22$ & $84$ & $25.6$ & $0.255$ & $0.019$ \\ 
                NB6200 & $7.70$ & $1.96$ & $0.20$ & $104$ & $25.3$ & $0.253$ & $0.019$ \\
                NB7320 & $7.88$ & $2.15$ & $0.23$ & $100$ & $27.1$ & $0.271$ & $0.020$ \\ 
                NB8025 & $8.14$ & $2.32$ & $0.34$ & $66$ & $28.2$ & $0.282$ & $0.022$ \\ \hline 
        \end{tabu}
        \vspace*{1mm}
        
        \small NOTES. --- Col. (1): NB filter. Cols. (2)--(4): Mean ($\overline{m}$), rms, and mean uncertainty ($\delta$) of all data points in the light curves in units of mJy. Col. (5): $\chi^2_\nu$ obtained by fitting a constant to the light curve. Col. (6): Intrinsic normalized variability measure, $\sigma_N = 100\sqrt{(rms^2-\delta^2)}/\overline{m}$. Cols. (7)--(8): Fractional variability amplitude and its uncertainty (\citealt{Rodriguez-Pascual1997,Edelson2002}).
\label{var}     
\end{table*}

\section{Time series analysis}\label{3}
The primary objective of this paper is to estimate the time delays between the NB passes located at 4250, 5700, 6200, 7320, and 8025\AA, which to a large extent trace the AGN continuum variations relatively free of contamination from the broad-emission lines. We used several methods to robustly determine the reverberation lags between the multiwavelength bands, as outlined below. All time lags are measured relative to the NB4250 light curve. A more detailed description of the methods (a) -- (e) can be found in \citet{Fian2022}.

(a) \textit{ICCF.}
A well-known method to estimate reverberation lags is the traditional interpolated cross-correlation function (ICCF) of \citet{Gaskell1986} and \citet{Gaskell1987}, as implemented by \citet{White1994}; see also review by \citet{Gaskell1994}. To properly perform cross-correlation function (CCF) analysis, uneven sampled light curves have to be interpolated. The time lag is determined by measuring the centroid of the points around the ICCF peak (above a certain threshold), and  
to estimate the errors of the inferred time lags we used the flux randomization and random subset selection (FR/RSS) method of \citet{Peterson1998,Peterson2004}. 

(b) \textit{ZDCF.}
One way to bypass interpolation is the use of a discrete correlation function (DCF; \citealt{Edelson1988}), which evaluates the correlation function in bins of time delay. In this work, we used the z-transformed discrete correlation function (ZDCF) of \citet{Alexander1997}, which applies Fisher's z transformation to the correlation coefficients. 
To measure the time delays between the multiband light curves, we took the centroid of the correlation function above 60\% or 80\% of the peak, and we estimated the errors using a maximum likelihood method that takes into account the uncertainty in the ZDCF points.

(c) \textit{Von Neumann Estimator.}
The von Neumann estimator (\citealt{VonNeumann1941,Chelouche2017}) does not depend on interpolation or binning of the light curves but is based on the regularity of randomness of the data. 
In this work, we used von Neumann's statistical estimator to find the relative time shift between two light curves that minimizes the level of randomness.

(d) \textit{JAVELIN.}
JAVELIN stands for\textit{\emph{ "Just Another Vehicle for Estimating Lags in Nuclei",}} and is a popular (parametric) Bayesian tool to measure reverberations lags (\citealt{Zu2011,Zu2013,Zu2016}). Instead of extracting peaks from empirical cross-correlation functions, it models the continuum variability of the quasar itself by assuming a damped random walk (DRW) process (\citealt{Kelly2009,MacLeod2010,MacLeod2012,Kozlowski2010,Kozlowski2016}). 
In this work, we used JAVELIN in spectroscopic and photometric RM mode. In the spectroscopic mode, we have two parameters for the continuum DRW model (amplitude and timescale of the quasar's stochastic variability) and three parameters for each lagging light curve (time delay, width of the smoothing function, and scaling factor). In photometric mode, JAVELIN models light curves in two different bands and estimates the contamination of the leading light curve to the longer wavelength band (additional parameter) and the corresponding time delay simultaneously.

(e) \textit{MICA.}
 MICA is a nonparametric method that determines the so-called transfer functions (see \citealt{Blandford1982}) for RM data, which reflect the structure of AGNs since the temporal behavior of spatially extended regions (outer parts of the accretion disk, BLR) is assumed to involve blurred echoes of the central ionizing continuum variations. 
The time lags are given by the first moment of the transfer functions, and the associated uncertainties are estimated as the standard deviation of the generated Markov chains.

(f) \textit{PRM.}
The photometric RM (PRM) developed by \citet{Chelouche2013} is a generalized approach to RM and is based on multivariate correlation analysis techniques. It is able to identify reverberation signals across the accretion disk and simultaneously identifies the relative contribution of an additional, slowly varying component (associated with the BLR) to the continuum signal. Observationally, neither the time lag nor the contribution of the BLR to the lagging continuum light curve is known. Those values are constrained by the requirement for a maximal Pearson correlation coefficient within the computational domain. A more detailed explanation of the performance of this method is given in \citet{Chelouche2013}.\\

We used 1000 Monte Carlo runs to obtain the lag distributions for methods (a)—(c) and (f). In the case of JAVELIN and MICA, we ran 10.000 Markov Chain Monte Carlo simulations. We applied a common time-lag search interval of $[\tau_{min},\tau_{max}] = [-15,10]$ days, and, for methods that required interpolation, we adopted a time step of 0.15 days.

\section{Results and discussion}\label{4}
In the subsequent section, we report the results of the multiband photometric study of the AGN MCG 08-11-011, including the derived continuum time lags between the different NBs, the corresponding lag spectrum, an estimate for the host-subtracted AGN luminosity, and the theoretical disk size as a function of luminosity.

\subsection{Continuum time lags}\label{41}
\begin{figure*}[!ht]
\centering
\includegraphics[width=15.4cm]{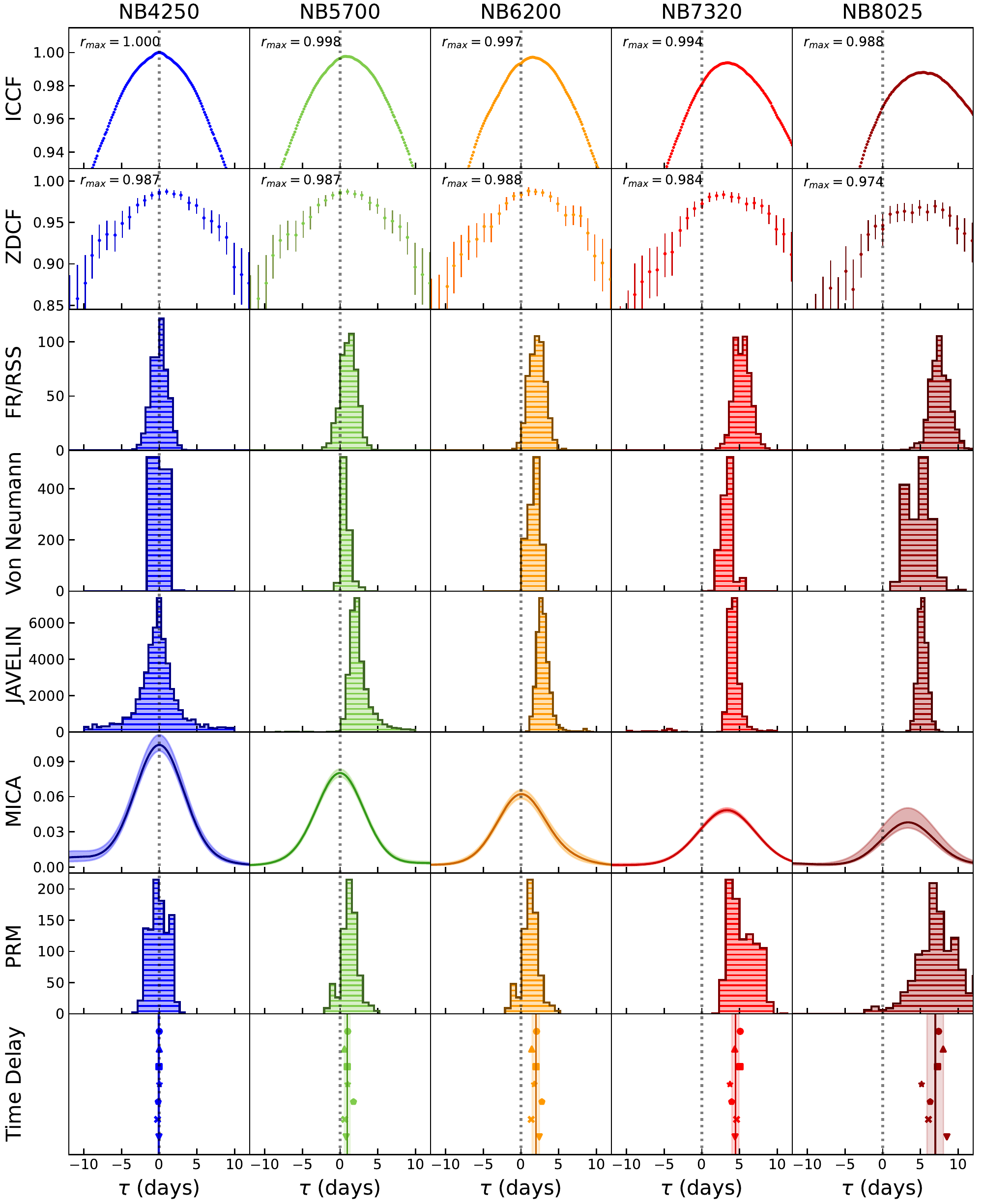}
\caption{From top to bottom: Partially interpolated CCFs, z-transformed DCFs, FR/RSS centroid distributions (for centroid $\ge 0.8\ r_{max}$), von Neuman estimator peak distributions, JAVELIN posterior distributions of lags (spectroscopic mode), MICA transfer functions, and PRM lag distributions for each NB relative to the 4250\AA\ band. In the bottom panel all time delays are plotted (in the same order as described before) on a vertical axis for illustration. The solid lines show the mean time delay of all methods together, and the shaded regions represent the corresponding standard deviation. These values are presented in Table \ref{list_td}.} 
\label{timedelays}
\end{figure*}

\begin{table}[h]
	\tabcolsep=0cm
	\renewcommand{\arraystretch}{1.35}
\caption{Summary of the time lags expressed in light days in the observer's frame between the five continuum light curves of MCG 08-11-011.}
\begin{tabu} {X[c]X[c]X[c]X[c]} 

\hline
\hline 
Method & Filter$^a$ & $\tau$ (days) & $r_{max}^b$ \\ \hline \hline
ICCF & NB5700 &  $1.0_{-1.0}^{+1.1}$  & $0.998$\\ 
 & NB6200 &  $2.1_{-1.1}^{+1.0}$ & $0.997$\\
 & NB7320 &  $5.1_{-0.8}^{+1.3}$ & $0.994$\\
 & NB8025 &  $7.4_{-1.1}^{+1.4}$ & $0.988$\\ \hline
ZDCF & NB5700 & $0.6_{-1.1}^{+1.0}$ & $0.987$\\
 & NB6200 & $1.4_{-0.6}^{+1.6}$ & $0.988$\\
 & NB7320 &  $4.4_{-1.4}^{+1.1}$ & $0.984$ \\
 & NB8025 & $8.0_{-3.5}^{+1.1}$ & $0.974$ \\ \hline
FR/RSS  & NB5700 & $1.0_{-1.1}^{+1.1}$ & --- \\
 & NB6200 & $2.0_{-1.1}^{+1.1}$ & --- \\
 & NB7320 & $5.0_{-1.1}^{+1.1}$ & ---\\
 & NB8025 & $7.3_{-1.3}^{+1.3}$ & --- \\ \hline
Von Neumann & NB5700 & $1.0_{-0.9}^{+0.5}$ & --- \\
\vspace*{-4mm}Estimator & NB6200 & $1.8_{-0.3}^{+0.6}$ & ---\\
 & NB7320 & $3.7_{-0.2}^{+0.2}$ & ---\\
 & NB8025 & $5.2_{-1.5}^{+2.4}$ & ---\\ \hline
JAVELIN & NB5700 & $1.3_{-0.3}^{+0.3}$ & ---\\
\vspace*{-4mm} (spectroscopic)& NB6200 & $2.7_{-0.3}^{+0.3}$ & ---\\
 & NB7320 & $3.8_{-0.4}^{+0.4}$ & ---\\
 & NB8025 & $5.3_{-0.6}^{+0.6}$ & --- \\ \hline
JAVELIN & NB5700 & $2.2_{-0.8}^{+1.6}$ & --- \\
\vspace*{-4mm}(photometric) & NB6200 & $2.8_{-0.7}^{+0.8}$ & --- \\
 & NB7320 & $4.1_{-0.6}^{+0.7}$ & ---\\
 & NB8025 & $7.3_{-1.2}^{+1.5}$ & --- \\ \hline
MICA & NB5700 & $0.6_{-0.5}^{+0.5}$ & --- \\ 
 & NB6200 & $1.4_{-0.5}^{+0.5}$ & --- \\ 
 & NB7320 & $4.6_{-0.6}^{+0.6}$ & --- \\
 & NB8025 & $6.1_{-1.6}^{+1.6}$ & --- \\ \hline
PRM & NB5700 & $0.8_{-0.6}^{+0.6}$ & --- \\
 & NB6200 & $2.4_{-0.5}^{+0.5}$ & --- \\ 
 & NB7320 & $4.4_{-1.0}^{+1.0}$ & --- \\
 & NB8025 & $8.5_{-2.5}^{+2.5}$ & --- \\ \hline \hline
Mean All & NB5700 & $1.0 \pm 0.5$ & --- \\ 
Methods & NB6200 & $2.0 \pm 0.6$ & --- \\ 
 & NB7320 & $4.5 \pm 0.6$ & --- \\ 
 & NB8025 & $7.1 \pm 1.1$ & --- \\ \hline \hline 
\end{tabu}
\label{list_td} 
\vspace*{1mm}
$^a\ $ relative to NB4250. \\
$^b\ $ maximum correlation coefficient.
\end{table}

We calculated the time lags (and their uncertainties) between the varying AGN continuum light curves at five different wavelengths using the various methods discussed in Section \ref{3}. All lags are measured with respect to the bluest (NB4250) light curve, and, for validation purposes, we also include the lag estimations of the 4250\AA\ NB relative to itself. In Figure \ref{timedelays}, we show the lag distributions or transfer functions for each light curve and method, and Table \ref{list_td} lists the corresponding lags and uncertainties. The last four rows give the overall mean time delay of all time-lag determination techniques, and the final uncertainties were obtained by estimating the standard deviation from the mean of all methods. From Figure \ref{timedelays}, we see that the lag distributions of the reference light curve relative to itself are symmetric and concentrated around zero as expected, while for the rest of the NB light curves the distributions are clearly shifted away from zero, and RM lags can be detected at high significance. 

For the ICCF, ZDCF, and FR/RSS, we computed the centroid time lags from all points above 60\% and 80\% of the peak value, leading to similar results. The lags derived from the ICCF, and ZDCF methods are consistent, indicating that the interpolation done in the ICCF does not introduce any artificial correlation. Also, the light-curve modeling techniques are able to capture reverberation lags, as can be seen for the JAVELIN and PRM posterior distributions, as well as for the MICA transfer functions. The lag distributions obtained from the von Neumann method after Monte Carlo simulation of FR/RSS as done for the ICCF analysis, yield similar results to those derived from the cross-correlation and light-curve modeling approaches. Thus, we find general agreements within uncertainties among the results of all methods used in this work. Combining all the lag estimates listed in Table \ref{list_td}, we obtain the mean time delays in the observer's frame relative to the 4250\AA\ NB, $\tau = 1.0 \pm 0.5$ days for NB5700, $\tau = 2.0\pm 0.6$ days for NB6200, $\tau = 4.5 \pm 0.6$ days for NB7320, and $\tau = 7.1 \pm 1.1$ days for NB8025. We note that in the case of the PRM, the contribution of the additional component is close to unity ($\sim$0.85) for all wavelength bands, thereby justifying the inclusion of the inferred lag estimates in the mean time delay calculation. Applying a weighted mean, we obtain slightly larger values for the time delays of the bluer wavelength bands ($\tau = 1.1$ days for NB5700 and $\tau = 2.2$ days for NB6200) and somewhat smaller values for the two reddest wavelength bands used in our RM campaign ($\tau = 3.9$ days for NB7320 and $\tau = 6.2$ days for NB8025); however, these are consistent (within errors) with the time lags obtained when using the ordinary mean.
\subsection{Lag spectrum}
\begin{figure*}[h]
\centering
\includegraphics[width=0.8\textwidth]{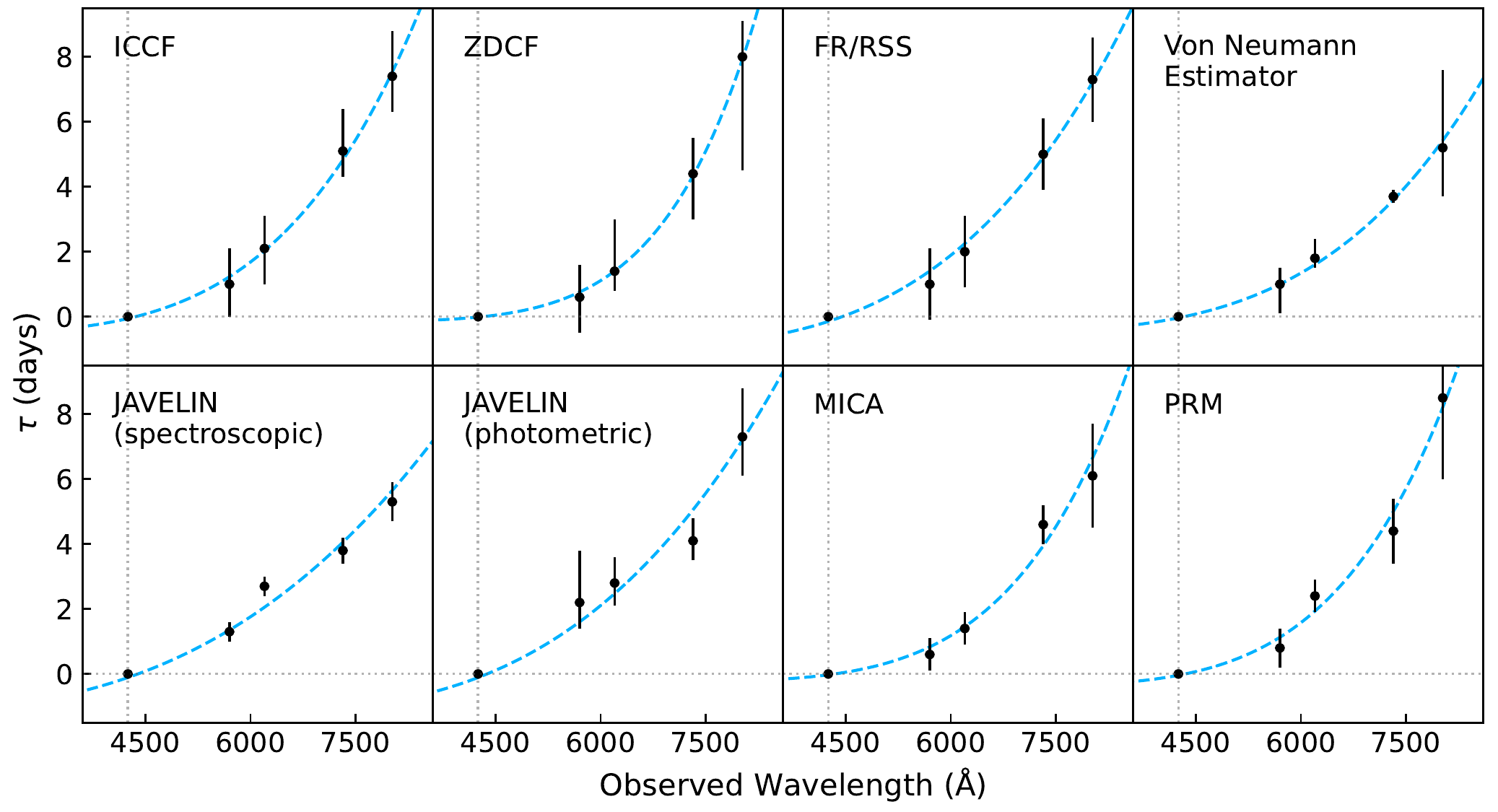}
\caption{Time lags (black circles) between multiband continuum light curves as a function of wavelength for the various methods described in Section \ref{3}. All lags are measured relative to variations at 4250\AA. The dashed blue lines show the best fit to the observed relation $\tau= \tau_0/\ [\left(\lambda/\lambda_0\right)^{\ \beta}-y_0]$ with $\tau_0$, $\beta$, and $y_0$ as free parameters (these values are presented in Table \ref{bestfits}).}
\label{r_vs_lambda_plot_extra}
\end{figure*}
\begin{figure}[h]
\centering
\includegraphics[width=8.9cm]{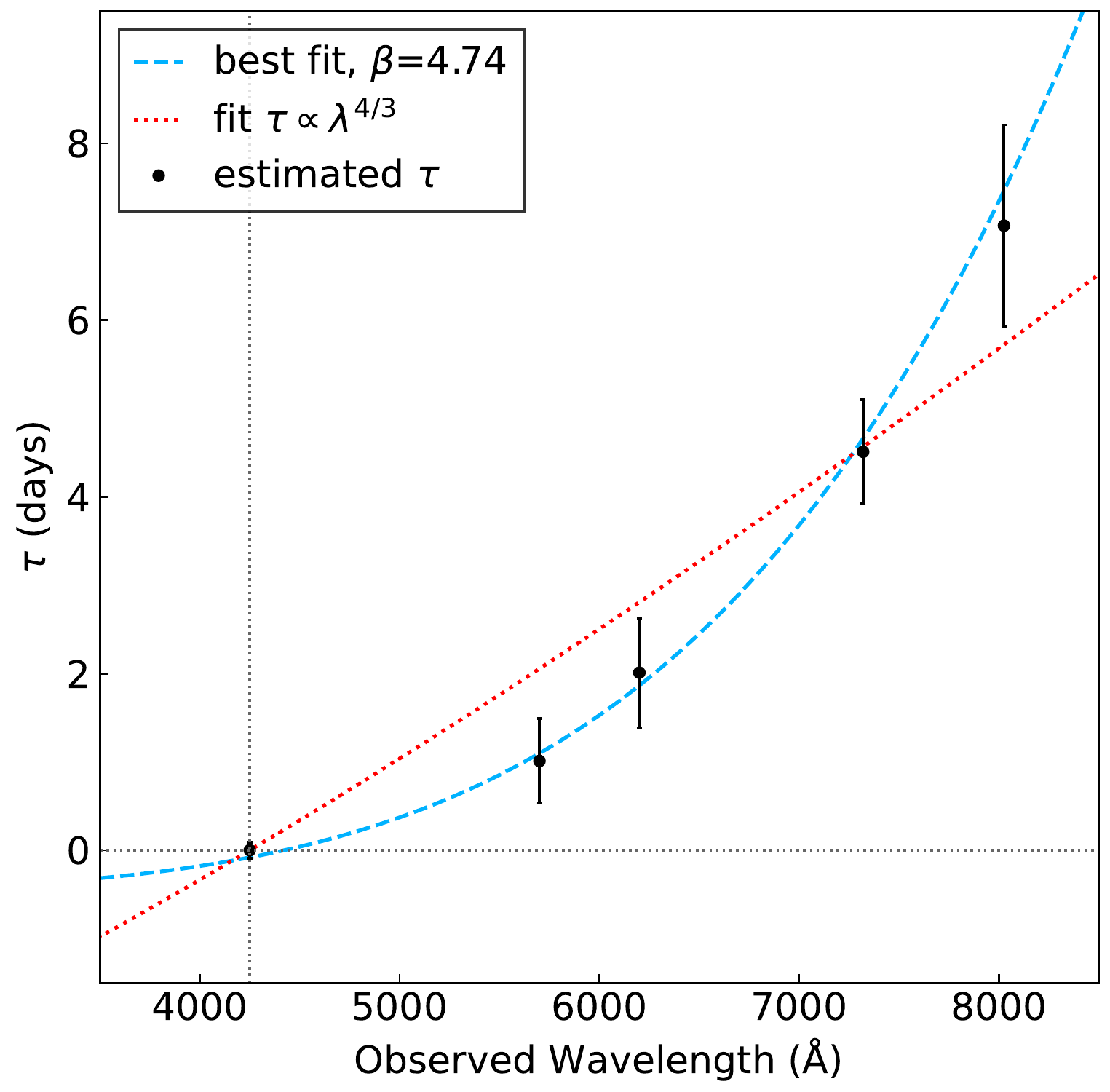}
\caption{Mean time lags (black circles) between multiband continuum light curves as a function of wavelength. All lags are measured relative to variations at 4250\AA. The dashed blue line shows the best fit to the observed relation $\tau= \tau_0/\ [\left(\lambda/\lambda_0\right)^{\ \beta}-y_0],$ with $\tau_0$, $\beta$, and $y_0$ as free parameters (these values are presented in Table \ref{bestfits}). The red dotted line is a fit with fixed theoretical power-law index $\beta = 4/3$, as expected for an optically thick and geometrically thin disk.}
\label{r_vs_lambda}
\end{figure}
Figure \ref{r_vs_lambda_plot_extra} displays the inter-band continuum time lags (relative to the 4250\AA\ light curve) as a function of wavelength for each of the methods discussed in Section \ref{3}. For a disk reprocessing model, we can translate the observed time-lag-wavelength relation to a wavelength-dependent emissivity profile, which in turn depends on the temperature profile of the accretion disk. To quantify this, we fit the observed continuum lags in the five different photometric NBs with a disk model of the following form:
\begin{equation}
    \tau = \tau_0 \left[ \left( \frac{\lambda}{\lambda_0} \right)^\beta - y_0 \right],
\end{equation}
where $\lambda$ is the observed wavelength, $\lambda_0$ is the reference band wavelength (here 4250\AA), and $\tau_0$, $\beta$, and $y_0$ are free parameters. The normalization $\tau_0 = R_{\lambda_0} / c$ measures the light crossing time across an accretion disk emitting at a reference wavelength, $\lambda_0$; the power-law index, $\beta,$ quantifies the temperature profile of the disk, $T \propto R^{-1/\beta}$; and $y_0$ allows the model lag at $\lambda_0$ to differ from 0. A list of best-fitting parameters is shown in Table \ref{bestfits}. From Figure \ref{r_vs_lambda} and Table \ref{bestfits}, we infer that the observed time lags, as well as the physical models, clearly favor a steeper slope than predicted by the standard disk temperature profile.
\begin{table*}[h]
    \tabcolsep=0cm
        \renewcommand{\arraystretch}{1.2}
        \caption{Best-fitting parameters to the inter-band lag spectra presented in Figure \ref{r_vs_lambda_plot_extra}.}
        \begin{tabu}{X[l]X[c]X[c]X[c]X[c]} 

                \hline
                \hline 
                Method & $\tau_0$ (days) & $\beta$ & $y_0$ & $\chi^2$\\  
                (1) & (2) & (3) & (4) & (5) \\ \hline
                ICCF & $0.48$ & $4.44$ & $1.12$ & $0.13$ \\
                ZDCF & $0.14$ & $6.38$ & $1.10$ & $0.03$ \\
                FR/RSS & $0.82$ & $3.62$ & $1.18$ & $0.20$ \\
                Von Neumann Estimator & $0.46$ & $4.02$ & $1.08$ & $0.48$ \\
                JAVELIN (spectroscopic) & $1.10$ & $2.88$ & $1.10$ & $5.42$ \\
                JAVELIN (photometric) & $1.10$ & $3.20$ & $1.10$ & $2.60$ \\
                MICA & $0.22$ & $5.42$ & $1.12$ & $1.55$ \\
                PRM & $0.34$ & $5.08$ & $1.12$ & $1.54$ \\ \hline
                All Methods together & $0.38$ & $4.74$ & $1.08$ & $0.17$ \\ \hline 
        \end{tabu}
                
                \small \vspace*{2mm}NOTES. --- Col. (1): Method (see Section \ref{3}). Cols. (2)--(4): Free parameters (scaling, slope, and shift). Col. (5): $\chi^2 = \sum\ (\tau_{obs}-\tau_{fit})^2 / \sigma_{obs}^2$.
\label{bestfits}        
\end{table*} 
In Figure \ref{r_vs_lambda}, we show the average time lag spectrum, and fit models with both $\tau_0$, $\beta$, and $y_0$ free to vary, as well as with $\beta$ fixed to $4/3$, corresponding to a standard thin accretion disk. The best fit yields $\beta = 4.74$ (dashed blue line in Figure \ref{r_vs_lambda}), resulting in a very steep lag-wavelength relation. The dotted red fit indicates that a disk reprocessing model with $\beta = 4/3$ cannot reproduce our data very well, contradicting the prediction for a geometrically thin disk with temperature profile of $T \propto R^{-3/4}$. In previous RM studies several authors observed similar trends in the lag spectra of AGNs (see, e.g., \citealt{Gaskell2007,Chelouche2019,Fian2022}) and attributed this to possible contamination by light being reprocessed from further away.

\subsection{Host-subtracted AGN luminosity}\label{host}
\begin{figure*}[h!]
    \centering
    \includegraphics[width=4.5cm]{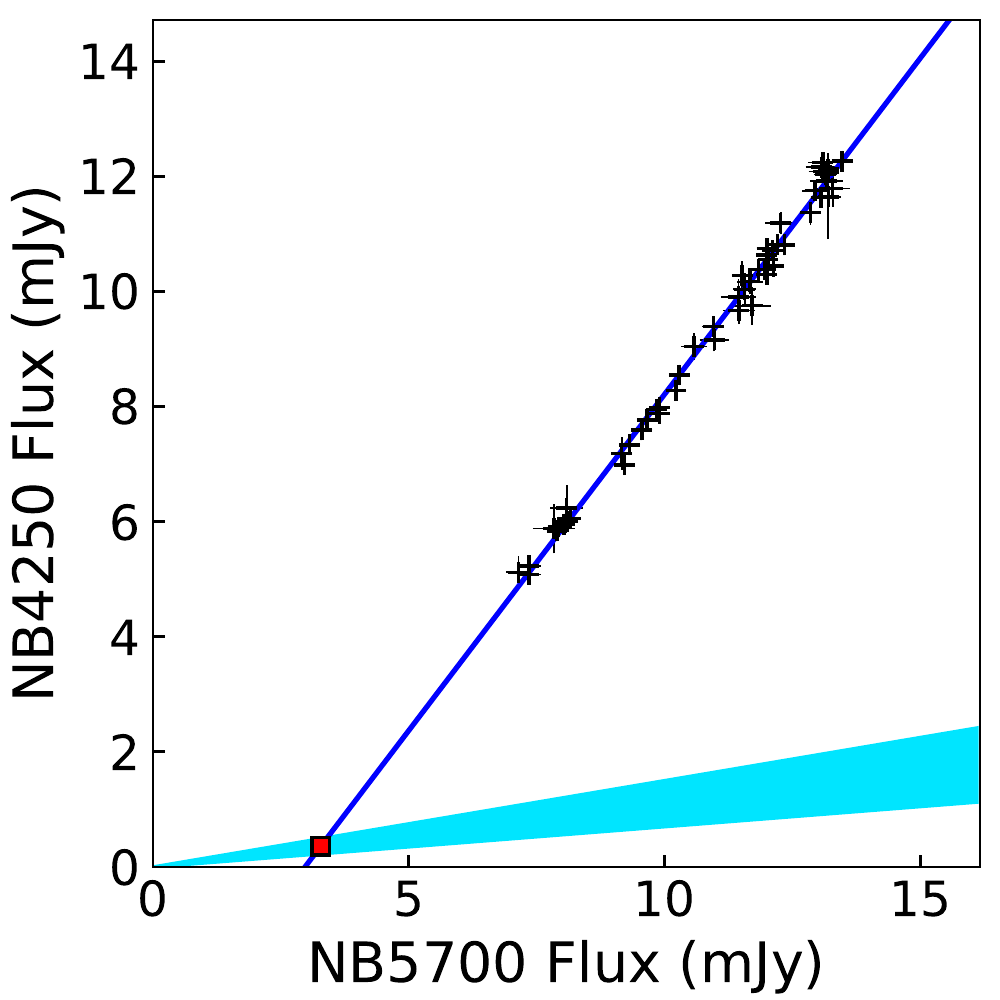} 
    \includegraphics[width=4.5cm]{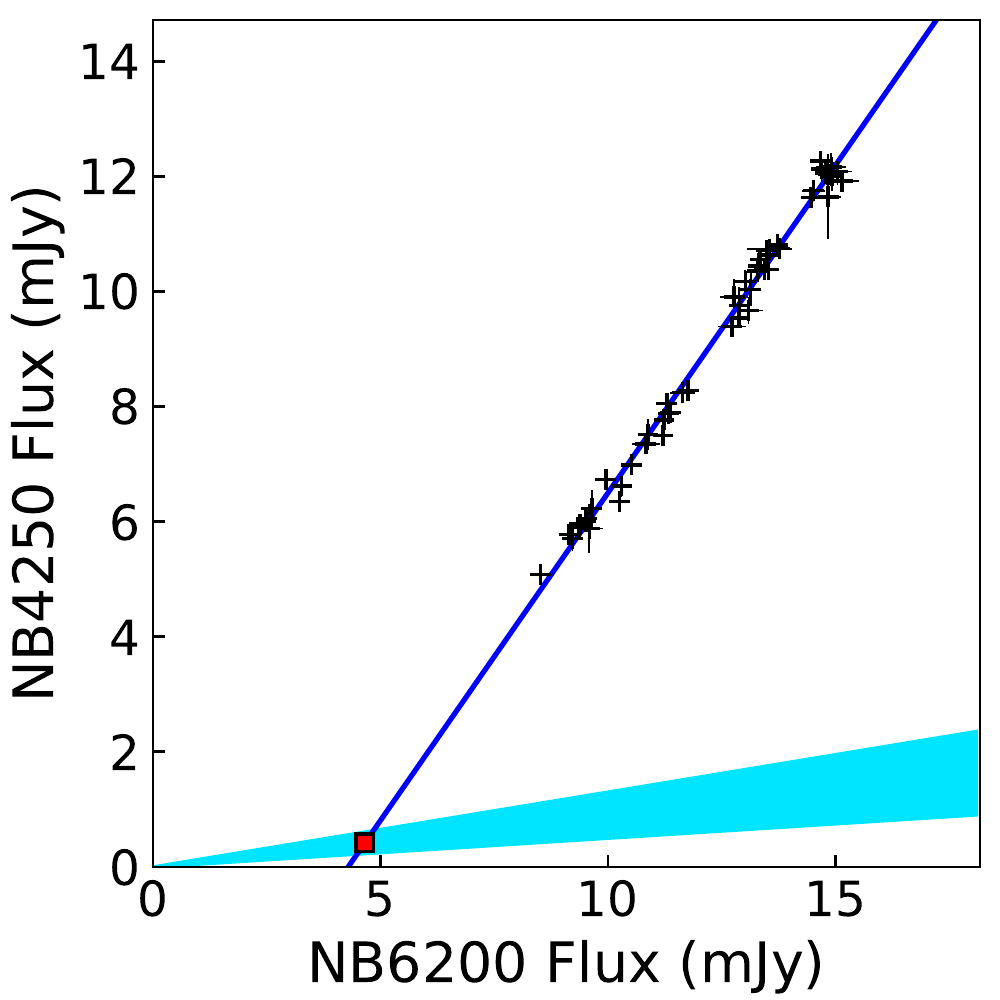}
    \includegraphics[width=4.5cm]{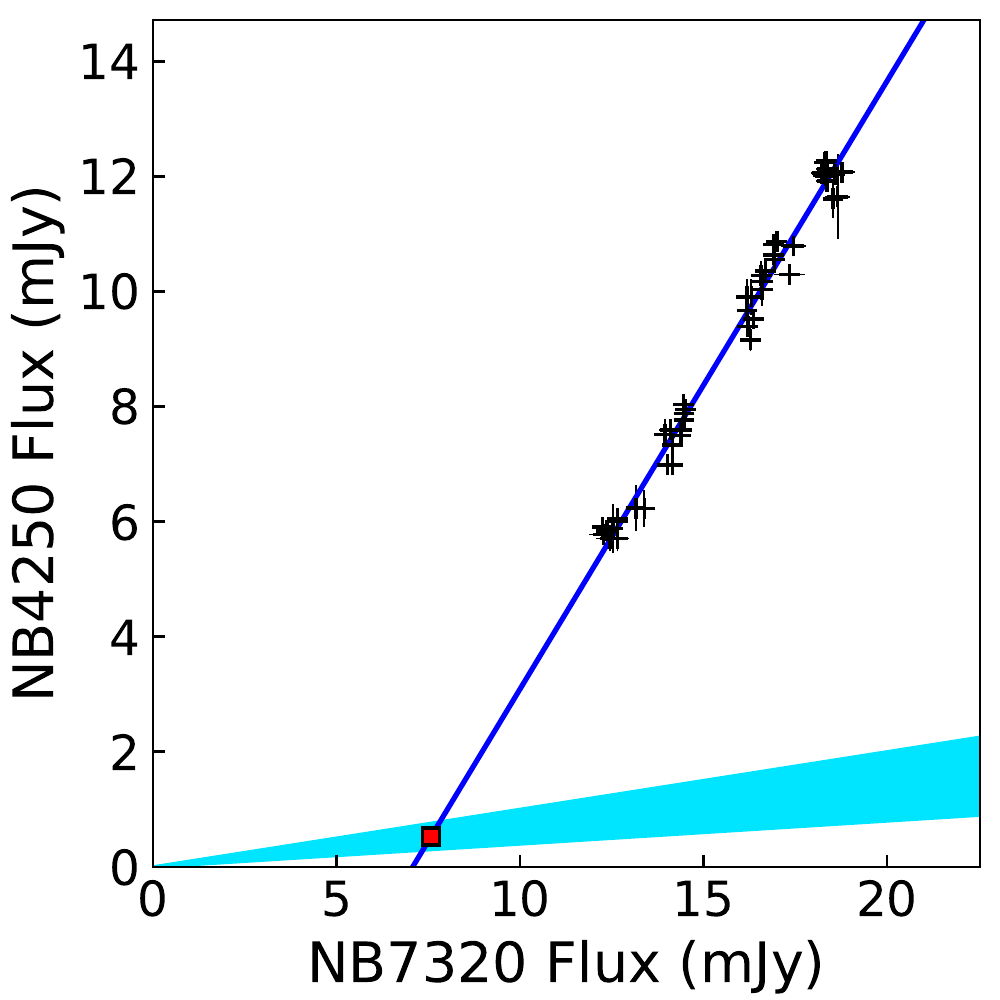} 
    \includegraphics[width=4.5cm]{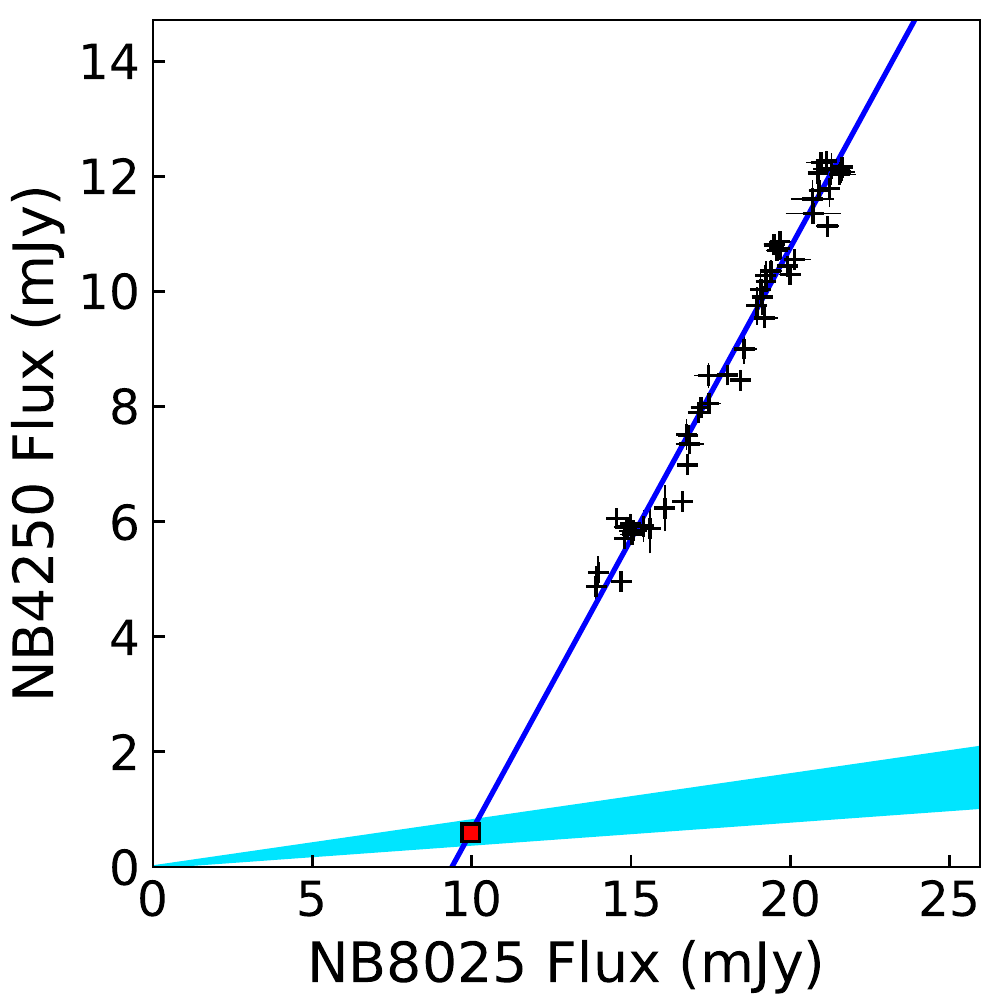}
    \caption{FVG diagram of MCG 08-11-011 between NB4250 and NB5700, NB6200, NB7320, and NB8025 (from left to right). Each data point is drawn as a thin cross in which the line length corresponds to the photometric uncertainties in the respective filters. A linear least-squares fit to the data points yields the AGN slope, plotted with the steep blue line. The cyan shaded area denotes the host color range from our multi-aperture photometry. The intersection between the AGN and the host galaxy slope gives the host contribution in the respective band within the aperture.}
    \label{FVG}
\end{figure*}

To determine the AGN's luminosity at a wavelength of 5100\AA, the contribution of the host galaxy to the nuclear flux has to be subtracted. To achieve this, we disentangled the constant host from the variable AGN flux inside our aperture by using the flux variation gradient (FVG) method originally proposed by \citet{Choloniewski1981} and further established by \citet{Winkler1992} and \citet{Sakata2010}. We plot data points for different filter pairs collected throughout the monitoring program in flux-flux diagrams in units of mJy (see Figure \ref{FVG}). As the observed source varies in luminosity, the fluxes in the FVG diagram will follow a linear relation with a slope (denoted by the symbol $\Gamma$; representing the AGN color) given by the host-free AGN continuum. The host, however, will show no variation. While the host slope passes through the origin, a linear least-squares fit to the data points yields the AGN slope. The intersection of the two slopes then allows us to determine the host flux contribution and to calculate the host-subtracted AGN luminosity at the time of the monitoring campaign - even without the need for high spatial resolution images (\citealt{Haas2011}). We note that the FVG diagrams were calculated taking into account the previously estimated time delays (in Section \ref{41}) between the different wavelength bands. The absolute flux calibration was carried out on the reference images (built from the individual NB frames) by comparison with the Pan-STARRS1 Catalog\footnote{https://catalogs.mast.stsci.edu/panstarrs/} (within a\ $20\arcmin$ distance of the target). Since the field is crowded, we obtained up to $\sim$150 comparison stars in the red filters. For each calibration star, we fit a black-body curve between the known $griz$ values and interpolate the flux to obtain the flux values for the central wavelengths of our NB filters. Finally, we calibrated and estimate the flux of MCG 08-11-011 in each NB and corrected the value for the Galactic foreground extinction (\citealt{Schlafly2011}).\\

Figure \ref{FVG} shows the NB4250 versus NB5700, NB4250 versus
NB6200, NB4250 versus NB7320, and NB4250 versus NB8025 fluxes of MCG 08-11-011. Linear least-squares fits to the flux variations in each NB filter pair yield $\Gamma_{AGN} = 1.18 \pm 0.02$ for NB4250 versus NB5700, $\Gamma_{AGN} = 1.15 \pm 0.02$ for NB4250 versus NB6200, $\Gamma_{AGN} = 1.10 \pm 0.02$ for NB4250 versus NB7320, and $\Gamma_{AGN} = 1.04 \pm 0.03$ for NB4250 versus NB8025. The host slope was determined by applying multi-aperture photometry on the stacked reference images as proposed by \citet{Winkler1992}. Fluxes measured at different apertures are used to infer the host galaxy color, and since the host galaxy contribution increases with the aperture, a linear fit between the fluxes approximates the host slope. We list the total (AGN + host) fluxes for each filter in Table \ref{fluxes} together with the mean host galaxy fluxes (obtained by averaging over the intersection area between the AGN and the host galaxy slopes) and the nuclear flux (calculated by subtracting the constant host galaxy component from the total flux). The listed uncertainties include the median errors of the calibration stars and errors caused by the black-body interpolation. The host contributes $\sim5$\% in NB4250, $\sim30$\% in NB5700, $\sim36$\% in NB6200, $\sim47$\% in NB7320, and $\sim54$\% in NB8025 to the total (AGN + host) observed fluxes. 

\begin{figure}
    \centering
    \includegraphics[width=8.8cm]{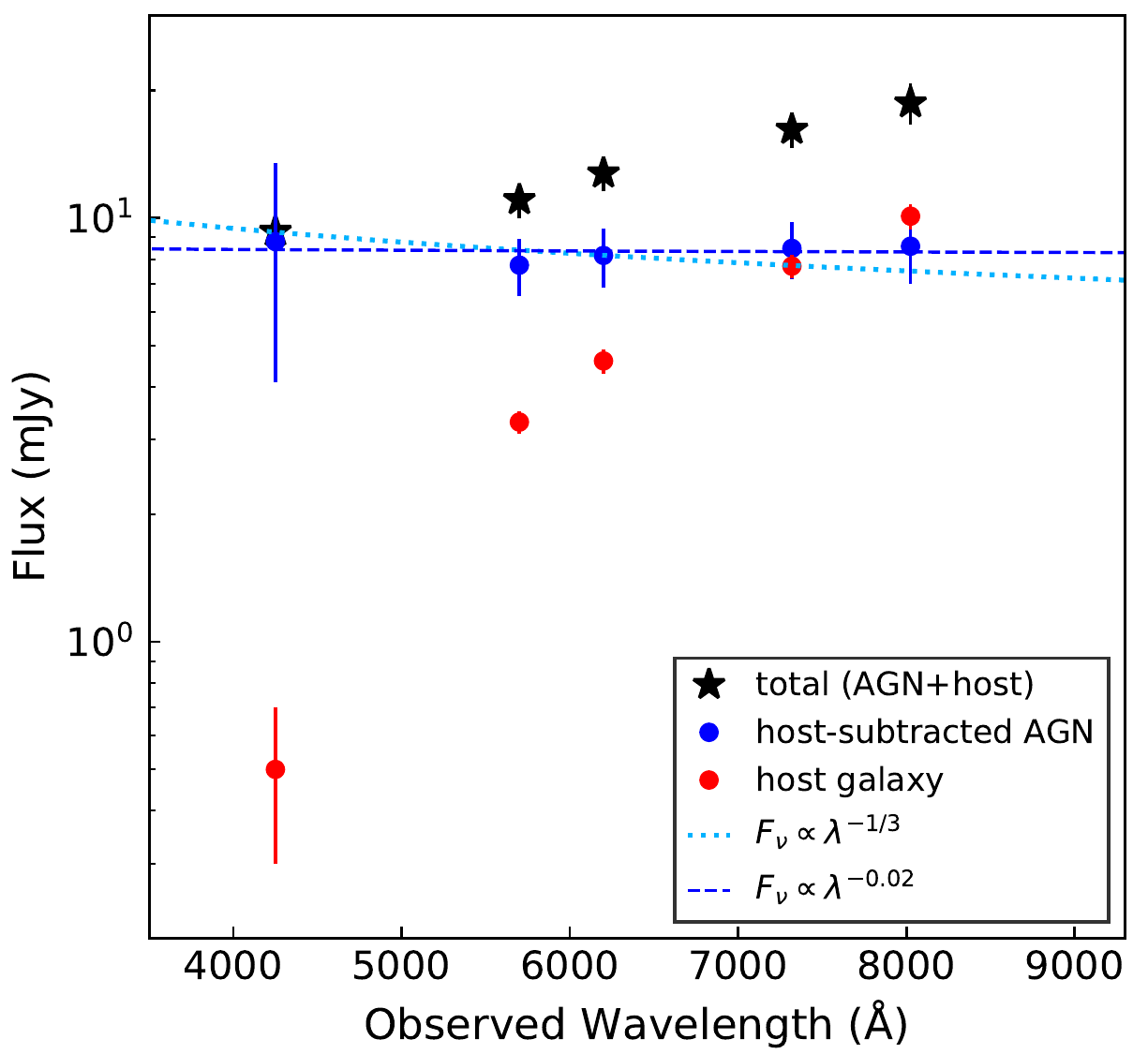}
    \caption{SED of MCG 08-11-011. Blue points show the host-subtracted AGN continuum with a power-law spectral shape of $F_\nu \propto \lambda^{-0.02\pm0.12}$ (dashed blue line). The dotted light blue line corresponds to a spectral shape as predicted by a standard Shakura-Sunyaev disk (with a spectral index of $1/3$).}
    \label{SED}
\end{figure}

\begin{table}[h]
\renewcommand{\arraystretch}{1.2}
\caption{Total (AGN + host), host galaxy, and AGN continuum fluxes.}
\begin{tabu} to 0.493\textwidth {X[c]X[c]X[c]X[c]} 

\hline
\hline 
Filter & Total (mJy) & Host (mJy) & AGN (mJy) \\
\hline 
NB4250 & $9.3\pm1.2$ & $0.5\pm0.2$ & $8.8\pm4.7$ \\ 
NB5700 & $11.0\pm1.0$ & $3.3\pm0.2$ & $7.7\pm1.2$ \\ 
NB6200 & $12.8\pm1.2$ & $4.6\pm0.3$ & $8.2\pm1.3$ \\ 
NB7320 & $16.2\pm1.6$ & $7.7\pm0.5$ & $8.5\pm1.3$ \\
NB8025 & $18.7\pm2.1$ & $10.1\pm0.7$ & $8.6\pm1.6$ \\ \hline 
\end{tabu}
\label{fluxes}    
\end{table}
To obtain the host-subtracted AGN flux of MCG 08-11-011 at a rest-frame of 5100\AA, we interpolated between the filters NB4250 and NB5700, adopting for the interpolation that the AGN has a power-law spectral shape ($F_\nu \propto \nu^\alpha$). At a distance of $D_L=93.10$ Mpc (\citealt{Yoshii2014}), this yields a host-subtracted AGN luminosity of $\lambda L_\lambda(5100\AA)=(4.21\pm0.65)\times 10^{43}$ erg s$^{-1}$.  The $\sim15$\% uncertainty includes the measurement errors, the uncertainty of the AGN and host slopes, and the AGN variations. In Figure \ref{SED}, we show the total (AGN + host) fluxes, the host-subtracted AGN fluxes, and the host fluxes as a function of wavelength. The power-law fit to the pure AGN fluxes yields $F_\nu\sim\lambda^{-\alpha}$, with $\alpha=0.02\pm0.12$, which is shallower than (but consistent within uncertainties with) the spectral index predicted by a standard Shakura-Sunyaev disk ($\alpha=1/3$). The host-subtracted RMS spectrum (values listed in Table \ref{var}) shows no spectral variation, which is accordant with the use of the Choloniewski diagrams. Thus, all fractional variability amplitude values are consistent with each other within their uncertainties.

It is worth mentioning that MCG 08-11-011 was previously monitored over four months in 2014 by \citet{Fausnaugh2017,Fausnaugh2018}, with the light curves spanning the broad-band \textit{ugriz} filters. Unlike our work, they observe a lag-spectrum consistent with geometrically thin accretion-disk models that predict a lag-wavelength relation of $\tau\propto\lambda^{4/3}$. They report significantly smaller lags (up to $\sim$2.6 days) than the ones inferred in the present paper using NB light curves, and they find that the disk is a factor of $3.3$ larger than predictions based on standard thin-disk theory. However, it is interesting to notice that \citet{Fausnaugh2017} estimated an (host-subtracted) AGN luminosity of $\lambda L_\lambda(5100\AA)\sim1.99\times 10^{43}$ erg s$^{-1}$, which is $\sim$2.1 times lower than our optical luminosity estimate. These differences in luminosity and measured reverberation lags indicate that the reprocessing may undergo changes on timescales of years. Up to now, only very few AGNs comprise high-cadence continuum RM data spanning timescales long enough to search for temporal changes in lags (one example is Mrk 110 which has shown evidence for a time-varying BLR contribution; \citealt{Vincentelli2021,Vincentelli2022}).

\subsection{Theoretical disk size}
A standard, geometrically thin, optically thick accretion disk radiates thermally and has a temperature profile of $T\sim R^{-3/4}$ (\citealt{Shakura1973}). The hot inner parts of the accretion disk emit in the UV ($\sim100-3000$\AA), while the cooler outer annuli emit in the optical and near-IR ($\sim3000-10000$\AA). As the short-wavelength emission from the X-ray-emitting corona and the inner edge of the disk varies, it irradiates the outer annuli and drives variations at longer wavelengths delayed by the light travel time across the disk (e.g., \citealt{Krolik1991}). Therefore, this model predicts (based on the object's SMBH mass and mass-accretion rate) theoretical time delays between short-wavelength and long-wavelength variations according to a given temperature-radius relation.

We compared the observed inter-band continuum lags with model predictions for thermal reprocessing following the method described by \citet{Fausnaugh2016} and \citet{Edelson2017}. Since the SMBH mass is highly uncertain, we substituted the product of the SMBH mass and mass-accretion rate with the target's optical luminosity, $L_{opt}$ (see Eq. (7) in \citealt{Davis2011}; for a detailed derivation, see \citealt{Fian2022}). Hence, we used the Shakura-Sunyaev model self-consistently and without the need to assume radiative efficiencies. The predicted light travel time $\tau$ (in days) relative to a reference time delay $\tau_0$ at a wavelength of $\lambda_0 = 5100$\AA\ can then be written as follows: 
\begin{equation}
(\tau-\tau_0) \simeq 2 \ \left( \frac{L_{opt}}{10^{45}\,\mathrm{ergs~s^{-1}}}\right)^{1/2}\times\ \left[\left(\frac{\lambda}{5100\,\mathrm{\AA}}\right)^{4/3}-1\right]\,\mathrm{days}.
\label{final_eq}
\end{equation}

We find that the inferred time lags are much larger (by a factor of $\sim 3-7$) than the theoretical lag estimates, which has been reported in previous works as well (\citealt{Jha2022,Fian2022,Montano2022,Edelson2019,Fausnaugh2018}). However, an accretion disk larger than predictions by a factor of $7.4$ for the longest optical wavelength band used in this work is striking since optical continuum RM campaigns typically find that continuum emission region sizes are $\sim 2-3$ times larger than expected from disk reprocessing models (\citealt{Cackett2022}). The discrepancy in MCG 08-11-011 is difficult to explain, and it is not clear that host contamination (even for an $\sim50\%$ contribution from the extended host galaxy to the observed PSF photometry light curve at that wavelength) and/or intrinsic reddening could fully account for the mismatch between theory and observations. One possible explanation is that AGN accretion disks are larger than model predictions and that their implied physics (e.g., the accretion disk temperature profile) is markedly different from that expected in the thin-disk scheme. Another possible explanation for the longer-than-expected continuum lags and their wavelength dependences is a substantial contribution of diffuse continuum emission from the BLR to the observed continuum signals and reverberation lags (e.g., \citealt{Cackett2018,Chelouche2019,Korista2019,Netzer2022}). Since we were not able to constrain higher moments of the transfer functions than the lags (i.e., the first moment), we could not test the pure accretion disk versus accretion disk-BLR origin for the time delays.

\section{Conclusions}\label{5}
We carried out photometric RM of the Seyfert 1 galaxy \mbox{MCG 08-11-011} using specially designed optical NB filters at the C18 telescope of the Wise Observatory, allowing us to trace the emission-line-free continuum at different wavelengths and measure inter-band continuum time lags. According to the disk-reprocessing \textit{lamppost} model (\citealt{Martocchia1996,Petrucci1997,Bao1998,Reynolds1999,Dabrowski2001}), photons arising from the innermost regions are reprocessed in the form of emission from the outer regions, resulting in a lag. The reverberation lags represent the light travel time across different regions of the disk and their trend with wavelength contains information about the disk's temperature profile. The high-cadence multi-wavelength observations at the Wise Observatory provide an excellent dataset to constrain inter-band reverberation lags efficiently. Our main results and conclusions are summarized below.
\begin{itemize}
    \item[(i)] All continuum light curves show significant correlated flux variations, which enabled us to carry out time series analysis to estimate the accretion disk size of MCG 08-11-011 using different cross-correlation and light-curve modeling methods. 
    \item[(ii)] We chose to measure lags relative to the NB4250 band as this is our bluest light curve, and we obtain mean time delays in the observer's frame of $\tau = 1.0 \pm 0.5$ days for NB5700, $\tau = 2.0\pm 0.6$ days for NB6200, $\tau = 4.5 \pm 0.6$ days for NB7320, and $\tau = 7.1 \pm 1.1$ days for NB8025. The inferred disk sizes are larger (by a factor of $\sim5$ on average) than predicted by the Shakura-Sunyaev accretion disk model, which is consistent with recent findings (\citealt{Jha2022,Fausnaugh2017,Fausnaugh2018,Fian2022,Pozo2019,Edelson2019,Cackett2018}). 
    \item[(iii)] The inter-band lags increase with wavelength, which provides strong evidence of disk reprocessing. However, the trend of lag versus wavelength does not match the $\tau \propto \lambda^{4/3}$ prediction of a standard geometrically thin disk. Phenomenological modeling shows that the data prefer a steeper lag-wavelength relation instead. This is in agreement with recent findings for Mrk 279, in which a diffuse continuum emission component was detected at the light curve level (\citealt{Chelouche2019}).
    \item[(iv)] A significant contribution of the host-galaxy was found in the reddest bands, and we estimated a monochromatic host-corrected AGN luminosity at 5100\AA\ of $(4.21\pm0.65)\times 10^{43}$ erg s$^{-1}$.
    \item[(v)] Interestingly, our results corroborate those from gravitational microlensing of strongly lensed quasars, which also find larger disk sizes than expected and a range of temperature profiles (\citealt{Jimenez2014,Motta2017,Fian2016,Fian2018,Fian2021,Cornachione2020a,Cornachione2020b,Cornachione2020c,Rojas2020}). While microlensing can only probe the continuum emitting regions in distant high-luminosity quasars, RM provides a complementary approach to investigate the accretion disk structure in low-luminosity AGNs.
\end{itemize}

Accretion disk sizes obtained through both gravitational microlensing and continuum RM indicate that the standard Shakura-Sunyaev disk assumption does not hold for the majority of AGNs studied so far, and it raises the question of the usage of the simple standard-disk model for AGN accretion disks. Thus, the discrepancy between theory and observations reinforces the suggestion that additional components (such as a contribution of the diffuse BLR continuum emission) may be needed while modeling the accretion disks in AGNs (\citealt{Jha2022,Montano2022,Vincentelli2021,Vincentelli2022,Fian2022}). Evidence of a non-disk component in the optical continuum of \mbox{Mrk 279} was reported by \citet{Chelouche2019}, indicating a possible explanation for the larger-than-expected continuum time lags. \citet{Vincentelli2021,Vincentelli2022} show, for the first time, that the BLR contribution may even vary in a single object, confirming the importance of considering the effect of emitting components different from the disk when studying the lag phenomenology in AGNs. Further multi-epoch observations over a broader range of wavelengths and a longer time baseline would be particularly valuable to search for evidence of diffuse continuum emission from the BLR and to better understand short-timescale variations in reprocessing behavior. Although mapping the entire accretion disk profile is only possible with intensive multiwavelength campaigns such as the AGN STORM (\citealt{Edelson2015,Fausnaugh2016,Kara2021}) with observations ranging from X-ray over UV/optical up to the far IR, the RM campaign at the Wise Observatory provides us with the opportunity to use well-sampled NB light curves free of prominent line emission to study inter-band continuum lags and to reach a more detailed understanding of their physical origin, albeit with a smaller wavelength coverage. This work can be extended to a larger sample of low-luminosity sources that are not accessible through microlensing, allowing us to further research the structure of AGN accretion disks and accretion mechanisms. 

\begin{acknowledgements}
We thank the anonymous referee for the constructive remarks on this manuscript. This work was financially supported by the DFG grant HA3555-14/1 and CH71-34-3 to Tel Aviv University and University of Haifa. This research also has been partly supported by the Israeli Science Foundation grant no. 2398/19. T. L. is supported by an appointment to the NASA Postdoctoral Program at Goddard Space Flight Center, administered by Oak Ridge Associated Universities under contract with NASA.
\end{acknowledgements}
\bibliographystyle{aa}
\bibliography{bibliography}

\end{document}